    \def\year{2021}\relax
\documentclass[letterpaper]{article} 
\usepackage{aaai21}  
\usepackage{times}  
\usepackage{helvet} 
\usepackage{courier}  
\usepackage[hyphens]{url}  
\usepackage{graphicx} 
\usepackage{natbib}  
\usepackage{caption} 
\usepackage{subfigure}
\usepackage{url}

\usepackage{soul} 
\usepackage{hyperref}
\usepackage{paralist}
\usepackage[dvipsnames]{xcolor}
\usepackage{multirow}
\usepackage[switch]{lineno}

\newcommand{\jiawen}[1]{\textcolor{Emerald}{ #1}}

\title{Modeling the Compatibility of Stem Tracks to Generate Music Mashups}

\author{

    Jiawen Huang\textsuperscript{*,2}, Ju-Chiang Wang\textsuperscript{1}, Jordan B. L. Smith\textsuperscript{1}, Xuchen Song\textsuperscript{1}, and Yuxuan Wang\textsuperscript{1}
}

\affiliations{

    \textsuperscript{\rm 1} ByteDance\\
    \textsuperscript{\rm 2} Queen Mary University of London
 
    jiawen.huang@qmul.ac.uk, \{ju-chiang.wang, jordan.smith, xuchen.song, wangyuxuan.11\}@bytedance.com

}

\begin{document}

\maketitle

\begin{abstract}

A music mashup combines audio elements from two or more songs to create a new work. To reduce the time and effort required to make them, researchers have developed algorithms that predict the compatibility of audio elements. Prior work has focused on mixing unaltered excerpts, but advances in source separation enable the creation of mashups from isolated stems (e.g., vocals, drums, bass, etc.). In this work, we take advantage of separated stems not just for creating mashups, but for training a model that predicts the mutual compatibility of groups of excerpts, using self-supervised and semi-supervised methods. Specifically, we first produce a random mashup creation pipeline that combines stem tracks obtained via source separation, with key and tempo automatically adjusted to match, since these are prerequisites for high-quality mashups. To train a model to predict compatibility, we use stem tracks obtained from the same song as positive examples, and random combinations of stems with key and/or tempo unadjusted as negative examples. To improve the model and use more data, we also train on ``average'' examples: random combinations with matching key and tempo, where we treat them as unlabeled data as their true compatibility is unknown. To determine whether the combined signal or the set of stem signals is more indicative of the quality of the result, we experiment on two model architectures and train them using semi-supervised learning technique. Finally, we conduct objective and subjective evaluations of the system, comparing them to a standard rule-based system.

\end{abstract}

\section{Introduction}
In Margaret Boden's account of how creativity works, ``combinational'' creativity---the juxtaposition of unrelated ideas---and ``exploratory'' creativity---the searching within the rules of a style for exciting possibilities---are two essential modes of creative thinking~\cite{boden2007creativity}.
Modeling these processes computationally is an important step for developing artificial creativity in the field of AI \cite{jordanous2014compcreativity}.
The combinatory possibilities and joy of exploration are perhaps two causes for the continued popularity of creating music mashups.

Mashups are a popular genre of music where new songs are created by combining audio excerpts (called \textit{samples}) of other songs. Typically, the vocal part from one song is juxtaposed with the instrumental part of another, 
although it is also common for basic mashups to include samples of 3 songs~\cite{boone2013mashing}.
However, creating mashups is challenging: it requires an expert's ear to decide whether two samples would make a good mashup, and it is time-consuming to search for pairs of songs that would work well together. Both issues are exacerbated when aiming to combine elements of three or more songs. 

Accordingly, efforts to assist users in creating mashups or to automate the process have continued for over a decade. 
At minimum, two samples being combined should have the same tempo and time signature, and they should not clash harmonically; these criteria informed early systems for assisting mashup creation~\cite{tokui2008massh,griffin2010beat}, but they are also easy to meet using beat-tracking, key estimation, and audio-stretching algorithms.
To predict the \textit{mashability} (i.e., the compatibility) of a candidate group of samples is more challenging, but allows one to automate the creation of mashups.

Previous methods for estimating compatibility have relied on rule-based systems with hand-crafted features~\cite{davies2014automashupper,lee2015automatic}, even though the criteria used by listeners to judge mashup quality are unknown and undoubtedly complex.
We thus propose to use a neural network to learn the criteria.
The central challenge in training such a model is that there is no training data: i.e., no datasets of audio samples with annotations defining which combinations of samples are better than others. To address this, we propose to use self-supervised learning (by leveraging existing music as training data) and a semi-supervised approach that maximizes the utility of the other data we create.

We create training data by applying a supervised music source separation (MSS) algorithm (e.g.,~\citealt{jansson2017singing,stoter20182018}) to extract independent stem tracks from existing music (i.e., separated vocal, bass, drum and other parts).
We can then recombine the stems to generate many new mashups, with the original combinations serving as ground truth examples of `good' mashups; in this way our model is \textbf{self-supervised}.\footnote{This may be different from conventional settings such as learning
a representation of signals or their temporal coherence
\cite{ misra2016shuffle,huang2018generating}.}
It is straightforward to use separated signals to help Music Information Retrieval (MIR) tasks such as music transcription \cite{pedersoli2020improving}, singing voice detection \cite{stoller2018jointly}, and modeling vocal features \cite{lee2019investigation}. However, to the best of our knowledge, no prior work has leveraged \textit{supervised} MSS for automatic generation of new music pieces.

The above explains how to acquire positive examples for training the model. To obtain negative examples, we can use random combinations of stems with different keys and tempo that are almost guaranteed to sound awful.
However, the extreme difference in compatibility between these two cases may lead to a highly polarized model that only regards stems as compatible if they were extracted from the same song.
To avoid this, we use \textbf{semi-supervised learning}: we create random mashups that meet the minimum requirements for viability---combinations where the tempo and key are automatically matched---and treat them as ``unlabeled'' instances.
This step aims to improve the reliability of the model, and it also means that our model sees more mashups, including many potentially ``creative'' ones, because the stems are sourced from different genres. This has been described as a key aspect of successful mashups: ``the combination of musical congruity and contextual incongruity''~\cite{brovig2012contextual}.

Our contributions can be summarized as follows. First, we propose a novel framework that leverages the power of MSS and machine learning to generate music mashups. Second, we propose techniques to generate data without human labels and develop two deep neural network architectures that are trained in a self- and semi-supervised way.
Third, we conduct objective and subjective evaluations, where we propose to use an unstudied dataset, Ayumix\footnote{About Ayu Creator Challenge 2020: \url{https://randomjpop.blogspot.com/2020/05/ayumi-hamasaki-launches-the-creator-challenge.html} accessed on March 10, 2021.}, to evaluate the task. The result demonstrates our AI mashups can achieve a good overall quality according to our participants.

\section{Related Work}

Early approaches to estimating mashability relied on fixed notions of what makes two sound clips mesh well.
AutoMashUpper (AMU) modeled the mashability of two clips as a weighted sum of harmonic compatibility, rhythmic compatibility and spectral balance, each of which is computed as a correlation between two beat-synchronous representations~\cite{davies2014automashupper}.
A system based on AMU was tailored to model good combinations of vocals and accompaniments, and included a constraint that the two parts should have contrasting amounts harmonic complexity, to avoid cases where both parts are complex or both are simple~\cite{lee2015automatic}.
Rule-based models of harmonic compatibility have been improved on~\cite{bernardes2017hierarchical} and deployed in mashup creation tools~\cite{maccas2018mixmash}, but all of these approaches share the potential weakness that a hand-crafted model of compatibility may fail to generalize, or to completely capture all of the subtle factors that contribute to balanced, high-quality mixes.
Nonetheless, AMU is a well-described, well-motivated baseline model that has been re-implemented for open-source use.


In contrast to AMU, our system uses a supervised model where the training data were obtained by running MSS on existing songs.
These steps were also taken to train Neural Loop Combiner (NLC), a neural network model that estimates audio compatibility of one-bar loops~\cite{chen2020neural}. 
However, NLC uses an unsupervised MSS algorithm designed to isolate looped content~\cite{smith2019unmixer}, resulting in a very different system to ours, which uses supervised MSS to isolate vocal, bass, drum, and other parts.

First, since vocals are looped less often, NLC likely had far fewer instances of vocals in its extracted training set. This is a drawback since vocals are an essential part of mashups.
Second, the data acquisition pipeline for NLC involves several heuristics to improve source separation quality, and the outputs are not guaranteed to contain distinct instruments (e.g., a positive training pair could include two drum loops).
In contrast, the supervised separation we use leads to highly distinct stems, which is more appropriate for creating mashups.
It also enables us to train a model where the input audio clips have a fixed role---namely, vocal, harmonic and drum parts---which is important, since the features that determine the compatibility likely depend strongly on the role. This design is also novel since it allows us to directly estimate the mashability of groups of stems instead of learning a representation space for embedding the stems, since mashability can be non-transitive.

A separate difference is that the authors of NLC chose to focus strictly on hip-hop, whereas our training dataset spans a wide variety of genres. We do this to obtain a more generalizable model (the semi-supervised technique explained later assists here too), and because much of the joy of mashups comes from the surprise of hearing two disparate samples from different genres work well together~\cite{brovig2012contextual,boone2013mashing}.

\section{Data Generation Pipeline} \label{sec:generate}

The pipeline aims to generate mashup candidates by mixing stems with different conditions. Then the candidates are sent to a machine learning model (described in the next section) to predict their compatibility.
The pipeline includes three modules, \textit{Music Source Separation} (MSS), \textit{Mashup Database} (MashupDB), and \textit{Mashup Generation}.


\begin{figure}[t]
      \centering
      \includegraphics[width =\columnwidth]{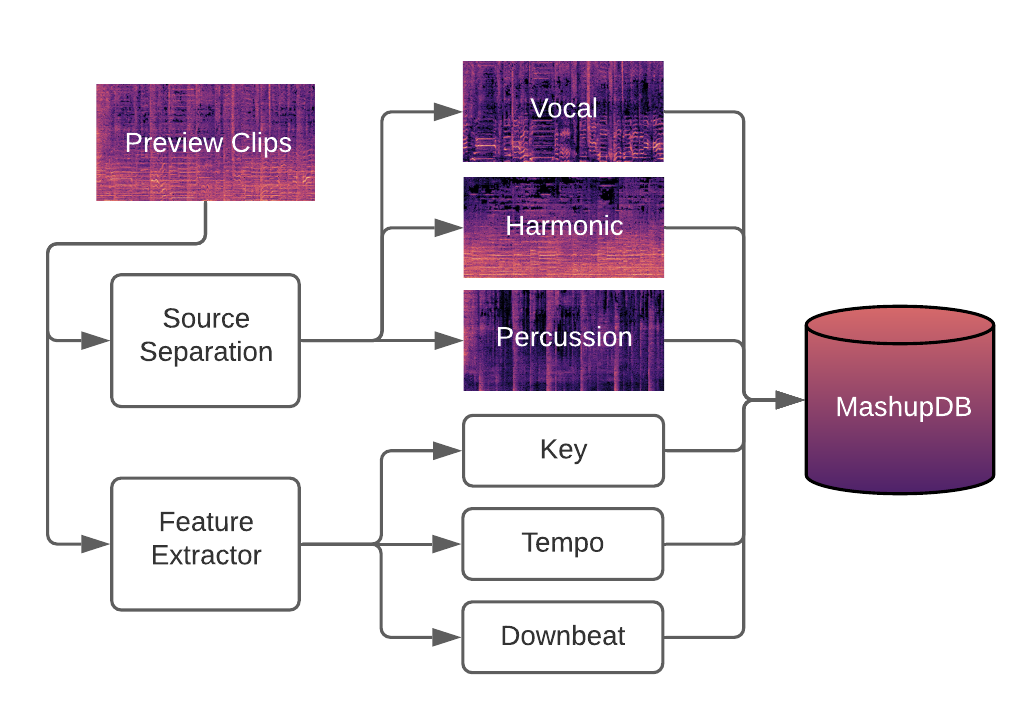}
      \caption{Diagram for MashupDB.}
      \label{fig:mashupdb}
\end{figure}

\begin{figure}[t]
      \centering
      \includegraphics[width =\columnwidth]{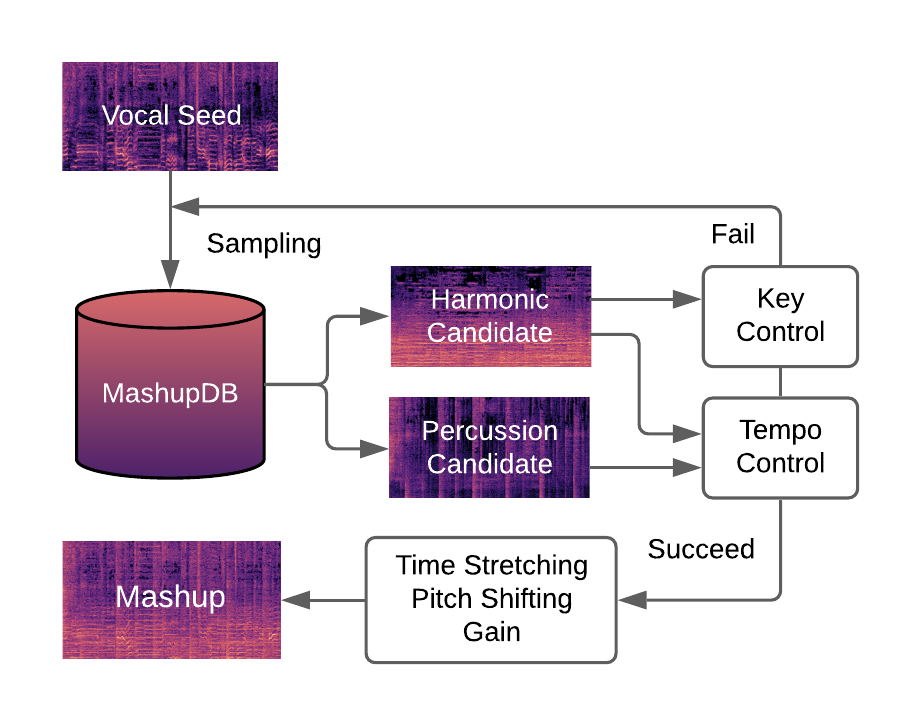}
      \caption{Mashup Generation pipeline.}
      \label{fig:remixer}
\end{figure}

\subsection{Music Source Separation}

We built our in-house MSS system based on a U-net encoder/decoder Convolutional Neural Network (CNN) architecture with skip connections \cite{jansson2017singing,pretet2019singing}. The network consists of 12 layers of encoders and decoders. Following the standard evaluation procedures on the MUSDB18 testset, our model achieved competitive Signal to Distortion Ratio (SDR) compared to the state-of-the-art \cite{defossez2019music}. 
Specifically, the mean SDR's for the four output stems, \textit{vocals}, \textit{drums}, \textit{bass}, and \textit{other}, are 7.21, 5.68, 5.51, and 3.74 dB, respectively. 
For reproducibility, we suggest one can use Spleeter \cite{hennequin2020spleeter}, an open-source tool with pre-trained models, as a replacement. It is expected to have similar effectiveness for generating the stem samples.

\subsection{Mashup Database} \label{sec:mashupdb}

To construct a MashupDB, we need a music database which ideally spans many genres.
To simplify, we use preview versions (clips around 30 seconds long) instead of the full audio tracks.
For each clip, we extract the stems using MSS (see Figure \ref{fig:mashupdb}), and we obtain the key, tempo, and downbeat information using madmom \cite{madmom}.
In this work, we consider three major stem 
types: vocal (\textit{vocals}), harmonic (\textit{bass + other}), and percussion (\textit{drums})---i.e., we mix the original \textit{bass} and \textit{other} stems for their similar purpose of representing the harmonic component of the accompaniment.


 
\subsection{Mashup Generation} \label{sec:remixer}

Figure \ref{fig:remixer} depicts the pipeline for generating mashups of different types. First, a vocal stem is randomly selected as the seed. The system then
searches in the MashupDB for harmonic and percussion stem candidates. Three conditions are allowed for the pipeline to generate mashups, namely \textit{original}, \textit{matched}, and \textit{unmatched} conditions. 

For \textit{original} condition, it selects the harmonic and percussion stems from the same clip of the vocal seed, and mixes them without adjusting the key and tempo. 

For \textit{matched} condition, the generated mashup shall satisfy basic harmonic and rhythmic requirements. With the vocal seed as the reference, this can be achieved by finding
the harmonic stem candidates within $\pm 3$ semitones in key,
and the percussion stem candidates having tempo within the ratio range of $[0.8, 1.2]$.
If no such harmonic and percussion stems exist, the pipeline will draw another vocal seed and start the process again. If multiple harmonic or percussion stems are found, it selects one at random and then adjusts the key and tempo respectively to match the vocal seed using Rubberband \cite{cannam2012rubber}. Like in~\cite{davies2014automashupper}, limiting the search space this way serves to avoid artefacts that can be caused by large amounts of pitch-shifting and time-stretching.
The resulting mashups have duration between 25 and 60 seconds.

The purpose of the \textit{unmatched} condition is to simulate cases where the key and tempo detection are wrong or the stems are incompatible.
We propose three strategies to this end.
First, \textit{unmatched key}, by disabling the key control and pitch shifting, the selected harmonic stem only satisfies the tempo constraint.
Second, \textit{unmatched tempo}, by disabling the tempo control and time stretching, it makes the three stems very likely to have clashing tempos. To increase the chance that this is immediately perceptible, the first downbeat of one stem is offset randomly by $\pm$ 1 second at most.
Third, \textit{unmatched key \& tempo}, by jointly applying the above two strategies, neither key nor tempo of the three stems meet the basic requirement.


For practical usage, we adopt the \textit{matched} condition to generate the mashup candidates for testing. To generate a good mashup, however, there are still numerous factors that the pipeline does not account for, such as chord, instrumentation, and groove. We believe those factors are too complicated to enumerate and justify manually. In addition, any errors in key, beat and downbeat detection and any artefacts from MSS and Rubberband can be propagated. All these factors can result in only part of the mashup candidates being good. To solve this problem, we make use of the pipeline to generate data to train machine learning models that could identify good pieces.


\section{Modeling the Mashability}

\subsection{Data Preparation}

To train the model to predict mashability, we build a training MashupDB based on a large music collection. Then, we employ the Mashup Generation pipeline to generate the positive, negative, and unlabeled data with the \textit{original}, \textit{unmatched}, and \textit{matched} conditions, respectively. 

Our system is fully audio-based, so for testing or practical use, one can build a new MashupDB different from the training one, depending on what musical sources the user wants to explore to generate mashups. 



The role of positive data is to guide the model to learn what a good combination of stems should sound like. 
Since the test samples are generated with the \textit{matched} condition, we also include the unlabeled data to ensure the model can see similar samples of the same condition, which we treat as the intermediate space between positive and negative data.

On the other hand, the role of negative data is to provide examples that do not meet the basic harmonic and rhythmic compatibility to be a musical piece. It has been shown that incorporating domain knowledge in designing the negative sampling strategies could help learning a robust model \cite{schroff2015facenet,wu2017sampling,riad2018sampling}.
For Neural Loop Combiner, Chen et al. investigated 5 ways of negative sampling, including random combinations of different loops, shifting or rearranging the order of beats~\cite{chen2020neural}.
Those techniques were devised in order to make 1-bar loops that already had the same tempo and duration incompatible. However, none of them were guaranteed to lead to incompatible pairs. The negative sampling strategies we have used are more clearly extreme---juxtaposing incompatible keys, tempos, and beat phases, in clips of 16 bars or longer---so most of their proposed strategies are not needed for our scenario.
Instead, manipulating the key and tempo for negative sampling is more likely reflecting the typical mistakes that an existing mashup system could make.


\subsection{Semi-Supervised Learning}

Based on the nature of the positive and negative data, perceptually distinguishing between them is very easy for humans.
We also observe similar trends that the training can achieve almost 100\% accuracy using only positive and negative labels on the validation set. 
To mitigate this concern, we explore semi-supervised learning methods that enable the model training to take the unlabeled data into consideration \cite{kingma2014semi,kipf2016semi}. Generally speaking, the quality of these unlabeled data is expected to vary widely, and thus we treat them as ``average'' examples with the hope that the training could help clarify them.


We adopt the method proposed by \cite{zheng2017unlabeled}, called \textit{label smoothing regularization for outliers} (LSRO). The original use case was on person re-identification in images, where the unlabeled images are generated by a generative adversarial network (GAN). 
The main idea of LSRO is to assign a uniform label distribution to the unlabeled data. In our case, supposing that the model output is two-dimensional classes (positive, negative), we assign a virtual label of $(0.5, 0.5)$ to each of the unlabeled data for loss computation. In this way, the learning process is regularized when an unlabeled example is close to either positive examples or negative examples in the feature space. 

As pointed out by \cite{zheng2017unlabeled}, LSRO exhibited superior performance to its semi-supervised learning counterparts, such as Pseudo Labeling \cite{lee2013pseudo} and All-in-One \cite{salimans2016improved}, in their application. In fact, we experimented with Pseudo Labeling, and it turned out not to be as good as LSRO in our pilot study. Therefore, we opt to present only LSRO in this paper.




\subsection{Network Architectures}


We formulate the task as a classification problem. The model can be described by $p(y \mid V, H, P)$, where $V$, $H$, and $P$ are respectively the input signals of the vocal, harmonic, and percussion stems, and $y \in \{0, 1\}$ is the binary label (good or bad). We take the output posterior probability as the mashability of the combination.

\begin{figure}[t]
      \centering
      \subfigure[ConvBlock]{\includegraphics[width=\columnwidth]{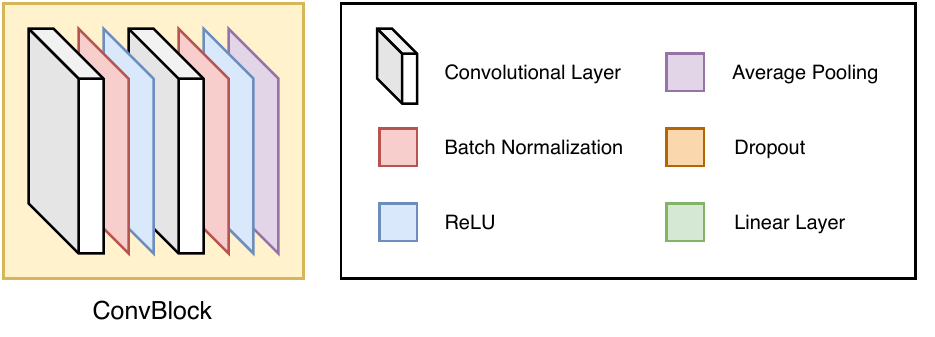}}
      \subfigure[PreMixNet]{\includegraphics[width=\columnwidth]{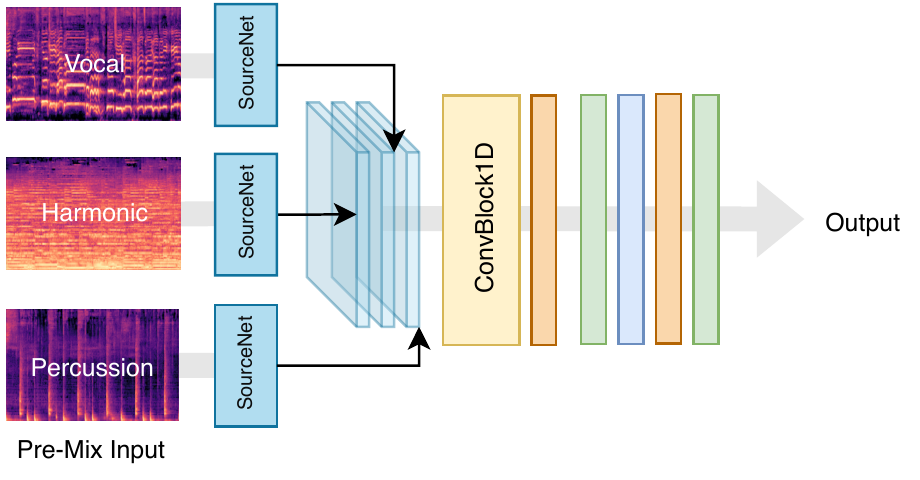}}
      \subfigure[PostMixNet]{\includegraphics[width=\columnwidth]{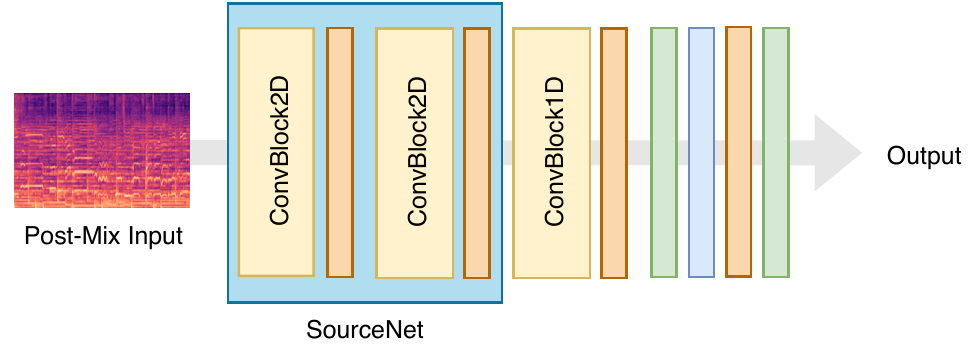}}
      \caption{The proposed model architectures.}
      \label{fig:network}
\end{figure}

We propose two types of model architecture, \textit{PreMixNet} and \textit{PostMixNet}, where the former takes the three individual stems as input (denoted as \textit{Pre-Mix}), and the latter takes the single downmixed track as input (denoted as \textit{Post-Mix}). The two models are illustrated in Figure \ref{fig:network}(b) and Figure \ref{fig:network}(c). 
When mixing audio tracks, it is common practice to 
judge the quality by listening alternately to 
the downmixed track and to each individual stem, or by comparing side-by-side the compatibility of a pair of individual stems \cite{senior2011mixing}. Because this process is complicated and different from person to person, we want to investigate whether the difference between the Pre-Mix and Post-Mix representations would affect the final performance. 


Figure \ref{fig:network}(a) defines the model components and the basic convolutional block (denoted as ConvBlock), which consists of two convolutional layers, each followed by a BatchNorm and a ReLU, and finally an average pooling layer. The convolutional layers in ConvBlock can be either 1D or 2D.
Both PreMixNet and PostMixNet take the 128-bin mel-spectrogram as input and have several convolutional blocks (Figures \ref{fig:network}(b) and \ref{fig:network}(c)). A 
unit we call
\textit{SourceNet}, which contains two ConvBlock2D's, takes the input of either \textit{Pre-Mix} or \textit{Post-Mix}. The purpose of SourceNet is to enhance the saliency of patterns on the mel-spectrograms \cite{mcfee2017structured}. We use a kernel size of (3,3) and a temporal pooling size of (2,1) for each ConvBlock2D.
Each SourceNet contains 4 convolutional layers (see \ref{fig:network}(c)) with 64, 64, 128, and 128 filters, respectively.
A ConvBlock1D follows to summarize the content for each frame. For a ~~ ConvBlock1D, we use kernel sizes of (3, 128) and (3, 1)
for the two convolutional layers, respectively (see \ref{fig:network}(a)), 
with 256 filters
and a temporal pooling size of (2, 1) for both. The time dimension is then reduced by average pooling, and the filter dimension is reduced by adding the maximum and average. Finally, two fully-connected layers (with 128 features) make the output.
In PreMixNet, we combine the three stem SourceNet outputs by concatenating along the filter dimension. The three SourceNets do not share the weights.

\section{Experiments}

Owing to people's different music backgrounds, tastes, and understandings of harmonic and rhythmic compatibility, judging the quality of a mashup is highly subjective. Following prior works \cite{davies2014automashupper,lee2015automatic,xing2020popmash}, we focus more on subjective evaluation but also provide the objective result for a reality check that the models are learning something useful.

\subsection{Datasets}
We built a training MashupDB based on an internal music collection of 33,192 music clips (with average duration about 30 seconds). It covers many genres of popular music, including Asian pop, Western pop, rock, folk, electronic, and hip-hop. Most of them have vocals, but a few are purely instrumental. We used the pipeline to generate a balanced dataset, denoted as \textit{in-house}, yielding a set of 51,507 examples, where 1/3 are positive, 1/3 negative, and 1/3 unlabeled. Note that any music collection could work as a training MashupDB as long as it is sufficiently large and generic.


For testing, we built two MashupDB's based on two publicly available datasets: Harmonix Set \cite{nieto2019harmonix} and Ayumix2020. We note that even though these two datasets may have few common songs to the training set, it is still valid because we did not use any human labels in training.

Harmonix Set contains 912 full-tracks covering genres in a wide range of popular western genres, such as pop, electronic, hip-hop, rock, country, and metal. Human annotations of beats, downbeats, and segments are available. To build the MashupDB, we extracted all the verse and chorus parts based on the segment labels, and used the beat and downbeat labels to obtain the tempo. There are in total 5,310 vocal seeds obtained from the verse and chorus segments of the 912 songs. In order to have a balanced test set, we applied the pipeline to generate, for each vocal seed, 5 clips with \textit{matched} condition, leading to 26,550 unlabeled mashup candidates.

The Ayumix2020 dataset originates from the Ayu Creator Challenge event,
in which the record label Avex
released 100 studio acapella tracks\footnote{\footnotesize{Studio acapella is the clean vocal stem track of a commercially released song. The data are available at } \scriptsize{\url{https://youtube.com/playlist?list=PL57sdSoJE6THHJyhWfFU7z1RdmWkdmL43}} \footnotesize{accessed on Mar 10, 2021.}} from J-pop star Ayumi Hamasaki. The aim was to encourage creators to create and share remixes using these tracks during the COVID-19 pandemic. We selected 30 Ayumix songs, segmented a chorus clip from each song, and extracted its corresponding key and tempo information from the original (non-acapella) version using madmom. We used Harmonix MashupDB to generate 180 unlabeled candidates for each song, meaning that the accompaniment stems (harmonic + percussion) to be combined with an Ayumix vocal stem are all from Harmonix Set. 


\begin{figure*}[!ht]
      \centering
      \includegraphics[width =\textwidth]{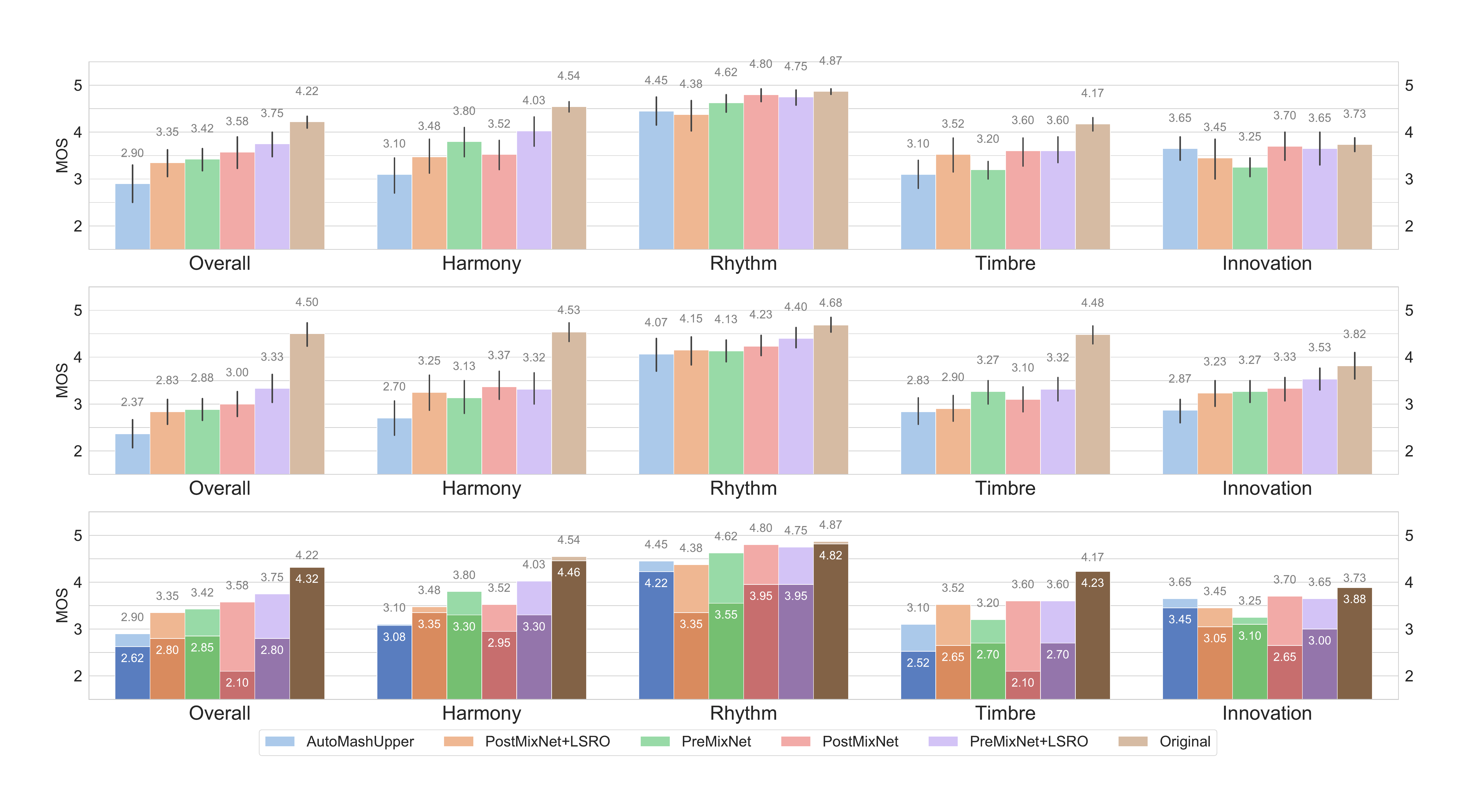}
      \caption{Comparison of different systems in subjective evaluation. All values are Mean Opinion Score (MOS). The upper and center sub-figures display the generation and retrieval tasks, respectively. The bottom sub-figure shows the MOS of top-20 mashups (in lighter colors) and bottom-20 mashups (in darker colors) for the retrieval task.}
      \label{fig:MOS}
\end{figure*}

\subsection{Model Training}
To compare the effectiveness of different model settings in the evaluation, we trained PreMixNet and PostMixNet with and without unlabeled data.
We used a ratio of $4\colon1$ for splitting the in-house dataset into training and validation sets. All models were trained with Adam Optimization with a $1\mathrm{e}^{-4}$ learning rate. 
All models were trained using an NVIDIA Tesla-V100 GPU for 3 days.



\subsection{Objective Evaluation}

\begin{table}[t]
\centering
  \begin{tabular}{l|c|c}
  \hline
                    & Accuracy     & Average Rank \\ \hline 
  PreMixNet         & {99.8\%}        & {1.0000}         \\
  PostMixNet        & {98.5\%}        & {1.0019}        \\
  PreMixNet + LSRO  & {98.9\%}        & {1.0024}        \\
  PostMixNet + LSRO  & {99.3\%}       & {1.0338}        \\ \hline
  \end{tabular}
  \caption{Results for objective evaluation.}
  \label{tab:obj}
\end{table}


For objective evaluation, we are interested in knowing  
(i) how well a trained model classifies the generated labeled data from the Harmonix Set (cross-dataset evaluation), and
(ii) how the positive data rank against the unlabeled data on the test set (accompaniment retrieval). To this end, we used the pipeline to further generate 5,310 negative examples with \textit{unmatched} condition and 5,310 positive examples with \textit{original} condition based on the vocal seeds.

The first objective can be evaluated in terms of accuracy on 
a binary prediction task
using 0.5 as the threshold. For the second objective, we used the test set of 26,550 unlabeled examples as the retrieval database. Following prior work \cite{chen2020neural}, we calculate the ranking position ($\geq$ 1) based on the probability scores for each positive example, and report the average rank.





As shown in Table \ref{tab:obj}, all proposed models achieve very high accuracy on labeled data and successfully rank the positive (original) one at the top. This demonstrates the model effectively discriminates between positive and negative examples, and between positive and unlabeled examples. However, this result does not show that the model is able to select unlabeled mashups that human audiences would prefer. Therefore, we conduct the following subjective evaluation.

\subsection{Subjective Evaluation}

\subsubsection{Baseline}
We compare our models with AutoMashUpper (AMU), which represents a rule-based automatic mashup creation system. We implemented the system based on an open-source tool.\footnote{\url{https://github.com/migperfer/AutoMashupper}}
AMU takes two stems as input. To get the mashability of a group of three stems, we take the average of the three pairwise mashabilities.

\subsubsection{Configuration}


Five systems were compared in the subjective evaluation (AMU and our four model variants), and we considered two evaluation tasks: \textit{generation} and \textit{retrieval}.


For the generation task, no vocal seed constraint is applied. Given a test MashupDB, we used the pipeline to generate as many \textit{matched} candidates as possible. Then, a model is employed to recommend the top candidates so that musicians may benefit from the ideas by listening to them. We used Harmonix Set alone for this task and picked the top 20 and bottom 20 mashups for each system in the listening test. This yielded  $(20+20)\times5=200$ samples to be evaluated.

For the retrieval task, we used the 30 selected songs from Ayumix as the test vocal seeds. We generated 180 mashup candidates for each vocal seed and selected the top 1 by each corresponding system, leading to 30 (queries) $\times 5$ (systems) $=150$ samples to be evaluated.


\subsubsection{Questionnaire}
Subjects were required to rate the quality of each test sample on a scale of 1 to 5 (1: awful, 2: poor, 3: fair, 4: good, and 5: excellent) according to five aspects: rhythm, harmony, timbre, innovation, and an overall score.
The questions were as follows:
\textit{Harmony}: How good is the harmonic compatibility between the vocals and music?
\textit{Rhythm}: How good is the rhythmic compatibility between the vocals and music?
\textit{Timbre}: Is each instrumental part clear and recognizable? How spectrally balanced and professional is the combination?
\textit{Innovation}: Does the combination sound unexpected without clashing?
\textit{Overall}: Does it sound like it was professionally made, not a randomly remixed piece?


\subsubsection{Methodology}
We developed a web-based platform for the listening test. Each page presented two audio samples which use the same vocal stem but different accompaniments. Subjects were not told about the relationship between the two samples, so they do not necessarily compare between them. Subjects were also asked to answer whether or not they have heard the vocal melody of the singer before. Pages were shown in a random order, and each page was rated by three subjects.
To assign the samples to web pages, separate strategies were used for the generation and retrieval tasks. For generation, each page presented a mashup (selected by a system) and its original version (generated by the pipeline with the \textit{original} condition), in a random order. This means that subjects were also required to rate the original version for a fair evaluation. For retrieval, because rating 6 samples (5 systems + 1 original) for a query vocal in one page is too much, we divided them into three pairs using a random permutation, so that each page still displayed two samples.


\subsubsection{Statistics of Subjects}
We recruited 40 subjects. Each subject rated at least 10 pages. According to self-report, 85\% said they listened to music on a daily basis, and 40\% had experience in music production.
Table \ref{tab:subject_stats} summarizes the statistics of the two tasks. 

\begin{table}[t]
\centering
  \begin{tabular}{l|cc}
  \hline
   & Generation   &  Retrieval  \\ \hline \hline 
  Number of subjects     &  28   & 12   \\        
  Time spent per page   &  95 sec  & 108 sec  \\
  Report vocal `never heard' &  68\%  & 75\%   \\ 
  Cronbach’s Alpha &  0.79  &  0.73  \\ 
  \hline
  \end{tabular}
  \caption{Statistics of subjects.}
  \label{tab:subject_stats}
\end{table}



 \subsubsection{Subjective Result}

Figure \ref{fig:MOS} shows the Mean Opinion Score (MOS) results of all cases. For each test clip, we take the median score from the three subjects, and average them over all clips for a system. In Tables \ref{tab:top_bot} and \ref{tab:PreMix_LSRO}, we also report the significance levels using Mann-Whitney U test for generation, and Wilcoxon rank-sum test for retrieval. In these tables, `O', `H', `R', `T', and `I' stand for overall, harmony, rhythm, timbre, and innovation, respectively.
Several points can be made as follows. 

Our proposed models outperform AMU by a large margin on `overall' and on harmony, and by a comfortable margin on timbre.
PreMixNet+LSRO performs the best among different variants in most of the cases, indicating that modeling individual stems and using unlabeled data with LSRO can improve predictions of mashability. The significance test results of PreMixNet+LSRO versus other systems are shown in Table \ref{tab:PreMix_LSRO}. 
Other than PreMixNet+LSRO, we see that PostMixNet also works well. Closer investigation reveals that PostMixNet prefers mashups with strong percussion stems, which are possibly easier to learn from the \textit{Post-Mix} positive examples alone. However, because it learns no unlabeled data, we can still find a few bad cases in the top 20 that degrade the MOS overall.

From Table \ref{tab:top_bot} we observe clearly that, for our models, the top-20 mashups outperform the bottom-20 ones significantly in most cases. By contrast, for AMU, this only happens on timbre. This demonstrates the effectiveness of a learned model in distinguishing between good and bad mashups. 

All the systems obtain higher MOS on rhythm. We attribute this to the straightforward beat structure of songs in Harmonix Set \cite{nieto2019harmonix}.
We also note that, by observing the low rhythm MOS of bottom-20, there are still many mashups having very poor rhythmic compatibility in the test set.
This is because tempo can vary within a clip, but the time-stretching by Rubberband is performed globally, leading to beat phase mismatches in the later measures.

The MOS for the original version can be regarded as the upper bound. But we note that these scores could be overrated, because in many cases when a subject is familiar with the song, it is very likely she/he would directly rate 5 for the original version without further judgement.
On the other hand, the original versions have noticeably fewer MSS artefacts, and this may also have influenced the ratings.

Finally, we find that the retrieval task is generally more difficult than the generation one. This is as expected because Ayumix vocals represent the most iconic J-pop style, while Harmonix Set is composed mostly by Western pop songs. We also ascribe the higher MOS of Ayumix's original versions to the relatively low quality of the generated Ayumix mashups, since it may be easier to find which is the original version by the subjects.

\begin{table}[t]
\centering
  \begin{tabular}{l|ccccc}
  \hline
  System  & O      & H      & R       & T        & I  \\ \hline \hline 
  Original          &  $\times$ & $\times$  & $\times$  & $\times$  & $\times$          \\ \hline
  AutoMashUpper     &  $\times$ & $\times$  & $\times$  & $.01$   & $\times$          \\                  
  PreMixNet         &  $.05$  & $\times$  & $.001$  & $.01$   & $\times$          \\
  PostMixNet        &  $.001$ & $.01$   & $.01$   & $.001$  & $.001$          \\
  PreMixNet + LSRO  &  $.001$ & $.05$   & $.01$   & $.01$   & $.05$           \\
  PostMixNet + LSRO &  $.05$  & $\times$  & $.01$   & $.01$   & $\times$          \\ \hline
  \end{tabular}
  \caption{Significance levels (p-value) between top-20 and bottom-20 mashups, where `$\times$' means no significance.}
  \label{tab:top_bot}
\end{table}

\begin{table}[t]
\centering
  \begin{tabular}{l|ccccc}
  \hline
  System (Generation) & O      & H      & R       & T        & I \\ \hline \hline 
  AutoMashUpper     &  $.01$  & $.001$  & $\times$  & $.05$   & $\times$    \\        
  PreMixNet         &  $.05$  & $\times$  & $\times$  & $.05$   & $.05$  \\
  PostMixNet        &  $\times$ & $.05$   & $\times$  & $\times$  & $\times$    \\
  PostMixNet + LSRO &  $.05$  & $.05$   & $.05$   & $\times$  & $\times$     \\ \hline
  \hline
  \hline
  System (Retrieval) & O      & H      & R       & T        & I \\ \hline \hline 
  AutoMashUpper     &  $.001$ & $.05$   & $\times$  & $.05$   & $.01$            \\     
  PreMixNet         &  $.05$  & $\times$  & $\times$  & $\times$  & $\times$            \\
  PostMixNet        &  $.05$  & $\times$  & $\times$  & $\times$  & $\times$            \\
  PostMixNet + LSRO &  $.01$  & $\times$  & $\times$  & $.01$   & $\times$            \\ \hline
  \end{tabular}
  \caption{Significance levels (p-value) on PreMixNet+LSRO versus other systems, where `$\times$' means no significance.}
  \label{tab:PreMix_LSRO}
\end{table}

\section{Conclusion and Future Work}

In this paper, we have proposed a novel framework to generate music mashups from separated stems and shown the potential of a learning-based approach to modeling the mashability. No human labels are needed, and the approach could work on any large music collection. The results of a listening test demonstrated the effectiveness of our models. Among the models we tested, adding unlabeled data to train a PreMixNet gave the best performance. 


Currently our system's efficiency is low because the model cannot predict the mashability before the key and tempo of each stem were adjusted. For improvement, we plan to learn an effective embedding that is invariant to key and tempo to hopefully generalize the mashability of stems using the metric learning techniques \cite{hu2014discriminative,lin2017automatic, humphrey2013feature, movshovitz2017no}. 
In addition, we plan to use the generated mashups as means of data augmentation to improve MIR tasks such as beat tracking and chord estimation.

\section{Acknowledgements}

We would like to thank Yi Deng and Yuan Wan for helping prepare the dataset, the participants for providing their ratings in the subjective evaluation, and the anonymous reviewers for their valuable comments.

\bibliography{aaai21}
\end{document}